\documentstyle[preprint,aps]{revtex}
\begin{document}
\def\br{\begin{eqnarray}}
\def\er{\end{eqnarray}}
\def\be{\begin{equation}}
\def\ee{\end{equation}}
\def\a{\alpha}
\def\D{\Delta}
\def\g{\gamma}
\def\G{\Gamma}
\def\l{\lambda}
\def\L{\Lambda}
\def\i{\int}
\def\m{\mu}
\def\n{\nu}
\def\({\left(}
\def\){\right)}
\def\s{\sigma}
\def\S{\Sigma}
\def\<{\left\langle}
\def\>{\right\rangle}
\def\gc{\<{\frac{\alpha_s}{\pi}}G^{\mu\nu}G_{\mu\nu}\>}
\draft
\title{
Squeezed gluon vacuum and the global colour model
of QCD
}
\author{
A.~A.~Natale~\footnotemark
\footnotetext{e-mail: natale@ift.unesp.br} 
}
\address{
Instituto de Fisica Te\'orica\\ 
Universidade Estadual Paulista\\
Rua Pamplona 145\\ 
01405-900, S\~ao Paulo, SP\\
Brazil}
\maketitle 
\begin{abstract}
We discuss how the vacuum model of Celenza and Shakin with a
squeezed gluon condensate can 
explain the existence of an infrared singular gluon propagator
frequently used in calculations within the global colour model. In particular,
it reproduces a recently proposed QCD-motivated model where
low energy chiral parameters were computed as a function of a
dynamically generated gluon mass. We show
how the strength of the confining interaction of this gluon propagator
and the value of the physical gluon
condensate may be connected.
\end{abstract}

\vskip 0.5cm
\newpage

\noindent

The Global Colour Model (CGM) of Quantum
Chromodynamics (QCD), whose main aspects have been reviewed
in the last years~\cite{tandy} -\cite{cahill},
is a quark-gluon quantum field
theory that very successfully models QCD for low energy
hadronic processes. In this approach an effective gluon
correlator models the interaction between quark currents,
and quark and gluon confinement may appear via the criterion
of absence of real $q^2$ poles for the propagators~\cite{pagels,munczek}.
This is, for instance, the case of an effective gluon propagator 
with an infrared singularity
like a delta function $\delta (k)$ at low energy~\cite{munczek}.
There are many recent calculations exemplifying the 
remarkable success of this procedure~\cite{rob,fr}. It relates the
hadronic properties to the Schwinger functions of quarks and
gluons, therefore, when comparing the theoretical calculations
to some low energy data, as pseudoscalars
masses and decay constants or other chiral parameters, we 
are learning how is the
infrared behavior of the quark and gluon propagators.
As the time goes on this semi phenomenological tool may 
reveal to be even more successful than the relativistic quark 
model or the bag model. However, the question for the mechanism which
leads to the infrared singularities present in the gluon propagator
in such calculations remains open.
 
The infrared enhancement of the gluon propagator due to the nonAbelian
character of the theory and in particular due to the gluon-gluon
self coupling, in principle could be understood in a rigorous study
of the QCD vacuum. Unfortunately the simple perturbative vacuum is
unstable~\cite{savvidy}, and there is no stable (gauge invariant)
coherent vacuum in Minkowski space~\cite{leut}. On the other hand
in the context of the construction of a gauge invariant, stable QCD
vacuum in Minkowski space, the squeezed condensate of gluons has
become a topic of interest to uncover the underlying dynamics
of the theory~\cite{celenza} -\cite{kogan}. Within this
class of vacuum models, 
Pavel {\it et al.}~\cite{pavel} recently proposed a phenomenological 
vacuum based on Abelian QCD which leads exactly to the infrared singularity
in the gluon propagator as considered in Ref.\cite{munczek}. On the
other hand calculations within the GCM suggest that the best fit
of chiral symmetry breaking parameters are obtained
with a propagator containing a delta function plus a propagator
that behaves as $1/k^2$ in the ultraviolet (consistent with
QCD) but damped at $k^2=0$~\cite{fr,nat}. Unfortunately, this
is not the case of Ref.~\cite{pavel} where a simple $1/k^2$ is
obtained together with the delta function. 

Even if a model of the squeezed QCD vacuum 
is not determined from first principles, its properties may be
very representative if they lead to a consistent phenomenology, 
indicating the path to the actual vacuum.
In this note we show
that the model of the QCD vacuum proposed by Celenza and
Shakin~\cite{celenza,celenza2} reproduces completely the
gluon propagator of Ref.~\cite{nat}, {\it i.e.} it gives a
singular part {\it \`a la} Munczek and Nemirovsky~\cite{munczek} as
well as a piece containing a dynamically generated 
mass~\cite{corn,bernard}. The effective dynamical gluon mass
is the unique parameter in the model,
and the gluon propagator obtained according to Ref.~\cite{celenza,celenza2}
is totally compatible with the
idea of the GCM. We will briefly discuss the CGM
and justify a gluon correlator which reproduces many aspects
of the chiral symmetry breaking phenomenology. Then, we show that this
gluon correlator naturally appears in the squeezed vacuum model of
Ref.~\cite{celenza,celenza2},
verifying that the parameters involved in the model (the strength of 
the confining interaction, the dynamical gluon mass and the value of 
the physical gluon condensate) are consistent with the ones in 
the literature. The coincidence between the models may indeed suggest
an interesting role for the Celenza and Shakin vacuum model. 

The action of the GCM can be obtained from the QCD generating
functional through the standard method presented in 
Ref.~\cite{tandy}:
\be
Z[\bar{\eta},\eta] = N \int D\bar{q}DqDA exp
\left( -S[\bar{q},q,A]+ \int d^4x (\bar{\eta}q+\bar{q}\eta) \right),
\label{e1}
\ee
where
\be
S[\bar{q},q,A]= \int d^4x \left( \bar{q} ( \not\!\partial + m
- \imath g \frac{\lambda^a}{2} \not\!A^a )q + \frac{1}{4}
G^a_{\mu\nu}G^a_{\mu\nu} \right),
\label{e2}
\ee
and $G^a_{\mu\nu} = \partial_\mu A^a_\nu - \partial_\nu A^a_\mu
+g f^{abc} A^b_\mu A^c_\nu$. In the above equation we have not written
the gauge fixing term, the ghost field term and its integration
measure. Introducing a source term for the gauge field and
writing
\be
exp(W[J]) = \int DA exp \left( - \int d^4x ( \frac{1}{4}
G^a_{\mu\nu}G^a_{\mu\nu} - J^a_\mu A^a_\mu ) \right),
\label{e3}
\ee
the generating functional becomes
\be
Z[\bar{\eta},\eta] = N \int D\bar{q}Dq 
exp( - \bar{q} ( \not\!\partial + m)q +\bar{\eta}q+\bar{q}\eta)
exp\left( W \left[ \imath g \bar{q}\frac{\lambda^a}{2} \gamma_\mu q \right]\right),
\label{e4}
\ee
where the spacetime integration is implied.
The functional $W[J]$ is the generator of connected gluon $n$-point functions
without quark-loop contributions. It may be written as
\be 
W[J] = \frac{1}{2} \int d^4xd^4y J^a_\mu (x) g^2D_{\mu\nu}(x-y)J^a_\nu (y)
+ W_R [J].
\label{e5}
\ee
The main characteristic of the GCM is the fact that we neglect the 
higher $n$-point functions contained in Eq.(\ref{e5}) and expressed
by $W_R [J]$. The effect of this approximation can only be measured in
the model building, but it is expected that the phenomenological
propagator $g^2 D_{\mu\nu}(x-y)$ retain most of the information 
about the non-Abelian character of QCD. With this approximation the
generating functional can be factorized as
\be
Z[\bar{\eta},\eta] = exp \left(  W_R \left[ \imath g \frac{\delta}
{\delta\eta} \frac{\lambda^a}{2} \gamma_\mu  \frac{\delta}
{\delta\bar{\eta}}\right]\right) Z_{GCM} [\bar{\eta},\eta].
\label{e6}
\ee
$Z_{GCM}$ is giving by
\be
Z_{GCM}[\bar{\eta},\eta] = N \int D\bar{q}Dq exp
\left( -S_{GCM}[\bar{q},q]+ \bar{\eta}q+\bar{q}\eta \right),
\label{e7}
\ee
with
\be
S_{GCM}[\bar{q},q] = \int d^4x \bar{q}(x) ( \not\!\partial + m)q(x) +
\frac{1}{2} \int d^4xd^4y J^a_\mu (x) g^2 D_{\mu\nu}(x-y)J^a_\nu (y).
\label{e8}
\ee
The action $S_{GCM}[\bar{q},q]$ together with the generating functional
$Z_{GCM}[\bar{\eta},\eta]$ defines the GCM. The idea of the model is
that the nonperturbative behavior that could be missed in the 
truncation performed above can be mostly represented by an effective
model in the infrared of
the gluon propagator in Eq.(\ref{e8}).

There are several discussions in the literature about the nonperturbative
behavior of the gluon propagator~\cite{robwil,mont,bernard}, as well as
ans\"{a}tze motivated by an impressive fitting of the low energy
QCD phenomenology. One such case is the propagator proposed by 
Frank and Roberts~\cite{fr} which yields
the expected QCD behavior in the ultraviolet and presents an integrable 
singularity at the origin. This is accomplished in Landau gauge by 
the following form (in the sequence all the momenta are in Euclidean space)
\be
g^2D_{\mu\nu}(k) = \left\{ \delta_{\mu\nu} -\frac{k_\mu k_\nu}{k^2}\right\}
\frac{g^2}{k^2[1+\Pi (k^2)]}\;,
\label{e9}
\ee
where,
\br
\Delta (k^2) &\equiv & \frac{g^2}{k^2[1+\Pi (k^2)]}
\nonumber \\
&=&  4\pi^2 d\left[ 4\pi^2 m_t^2\delta^4(k) +
\frac{1-e^{(-k^2/[4m_t^2])}}{k^2}\right]\;,
\label{e10}
\er
with $d=12/(33-2n_f)$, and $n_f=3$ (considering only three quark flavors).  
The mass scale $m_t$ determined in Ref.~\cite{fr} was interpreted
as marking the transition between the perturbative and
nonperturbative domains. We recently modified the above gluon propagator in
order to corroborate with the suggestion of several 
works~\cite{corn,bernard} that the gluon can 
develop a dynamical mass and
established that the 
unique parameter, $m_t$, could conveniently be substituted by the scale 
associated to this mass. Then, the Eq.(\ref{e10}) should be 
modified to~\cite{nat}
\br
\Delta (k^2) &\equiv & \frac{g^2}{k^2[1+\Pi (k^2)]}
\nonumber \\
&=&  4\pi^2 d\left[ 4\pi^2 m_g^2\delta^4(k) +
\frac{1}{k^2 + m^2(k^2)}\right],
\label{e11}
\er
where, 
\be
m^2(k^2) = m_g^2\frac{m_g^2}{k^2+m_g^2}\;,
\label{e12}
\ee
is the dynamical gluon mass. Eq.(\ref{e12}) interpolates between the large 
momenta behavior of the gluon mass predicted by 
the operator product expansion~\cite{lavelle}
\be
m^2_g (k^2) \sim  \frac{34 N \pi^2}{9(N^2-1)}
\frac{\gc}{k^2},
\label{e13}
\ee
and its finite constant infrared behavior as discussed in Ref.~\cite{corn}.
In Eq.(\ref{e13}) $\gc$ is the gluon condensate.
Eq.(\ref{e11}) has in its first term the singularity of Eq.(\ref{e10}), 
and the dynamical gluon mass gives a natural (and needed) damping of 
the second term at small $k^2$. 
In Ref.~\cite{nat} we obtained
a good fit to some chiral parameters ( as the pion mass and the
quark condensate) using the value $m_g \sim \, 600
\,\, MeV$. Our next step is to show how the model of
Ref.~\cite{celenza,celenza2} can fully describe the behavior of Eq.(\ref{e11}).

Colour-singlet coherent states can be formed integrating over the group
elements associated with the colour group
\be
|{\bf Z}_o(t)> = N \int [dg] U(g) |{\bf Z}(t)>,
\label{e14}
\ee
where $N$ is a numerical factor, $g$ specify an element of the group
$SU(N)$ and $U(g)$ are rotation operators whose properties are described
in Ref.\cite{celenza2}. Celenza {\it et al.}~\cite{celenza2} were able
to form a squeezed state $|{\bf Z}_o>$ such that
\be
<{\bf Z}_o|:{\bf E}_a({\bf r},t)\cdot {\bf E}_a({\bf r},t):|{\bf Z}_o> = 0,
\label{e15}
\ee
and
\be
\frac{g^2}{2} <{\bf Z}_o|:{\bf B}_a({\bf r},t)\cdot 
{\bf B}_a({\bf r},t):|{\bf Z}_o> \neq 0. 
\label{e16}
\ee
Meaning that the condensate is purely magnetic (${\bf B}_a({\bf r},t)$), 
while the color-electric
field (${\bf E}_a({\bf r},t)$) leads to a vanishing condensate. 

Many of the properties
of this model were obtained previously with the assumption that the
gluon field could be decomposed into a constant condensate field 
${\cal G}_{\mu}^a$ and the quantum fluctuations $g_{\mu}^a(x)$ around
it~\cite{celenza},
\be 
G_{\mu}^a(x) = {\cal G}_{\mu}^a + g_{\mu}^a(x).
\label{e17}
\ee
Taking into account Eq.(\ref{e17}) and $<g_{\mu}^a(x)>=0 $, one obtains the
decomposition of the nonperturbative gluon propagator into two parts,
\be
g^2 G_{\mu\nu}^{ab} (x-y) = <g^2 {\cal G}_{\mu}^a{\cal G}_{\nu}^b >
+ <g^2 g_{\mu}^a(x) g_{\nu}^b(y) > .
\label{e18}
\ee
Note that the choice of Eq.(\ref{e17}) is a purely phenomenological one,
and the correct construction of the model goes through the steps of
Ref.\cite{celenza2}. However, in this way it is easier to verify some
of the consequences of the model. First, it does generate a massive
effective lagrangian~\cite{celenza}
\be
{\cal L}_{m_g} (x) = - \frac{1}{4} G_{\mu\nu}^a (x)G^{\mu\nu}_b (x)
+ \frac{m_g^2}{2} g^{\mu}_a(x) g_{\mu}^a(x) + ...
\label{e19}
\ee
Secondly, the gluon mass $m_g^2$ is equal to $(15/32)<g^2{\cal G}^2>$. In
Ref.~\cite{celenza} the square of this condensate has been identified with the
phenomenological gluon condensate of Shifman {\it et al.}~\cite{shif}, but we
do not pursue this point since the gluon mass may contain a large part of
dynamics, and we do not see any {\it a priori} reason for the identity of the
condensates (although it looks phenomenologically plausible). Finally, what
more interest us is that in Euclidean momentum space the effective
nonperturbative gluon propagator corresponding to the decomposition of
Eq.(\ref{e18}) has the form~\cite{shakin} 
\be
g^2\Delta (k^2) \propto \left[ 16 \pi^4 m_g^2 \delta^4(k) +
\frac{1}{k^2 + m_g^2}\right],
\label{e20}
\ee
where the
$\delta^4(k)$ comes from the condensate field and the massive propagator is
a consequence of both fields (since the condensate field generates
the gluon mass).
Once we have not saturated the condensate resulting from Eq.(\ref{e18}) with
the one of Shifman {\it et al.}, the value of the dynamical gluon mass can be
considered a free parameter and not yet related to the gluon condensate.
Note that we do
have some freedom in the definition of the factors of
Eq.(\ref{e11})~\cite{nat}, as well as in the definition of the gluon mass in
Eq.(\ref{e20}), this is why we may affirm that Eq.(\ref{e11}) and
Eq.(\ref{e20}) are totally compatible. The important point is that
the form of the propagator in Eq.(\ref{e20}) is
consistent with the propagator of Ref.~\cite{nat}, and
gives a meaning for the confining propagators of the GCM. It is
interesting that the proposal of Ref.~\cite{nat} was to modify the propagator
of Ref.~\cite{fr} introducing the concept of the dynamical gluon mass, and
it turned out that this modification just led to the model of
Ref.~\cite{celenza}.

In the Celenza and Shakin model the gluon mass and the strength of the
confining interaction are related to
the constant condensate field, and the square of this one was identified
with the gluon condensate of Shifman {\it et al.}~\cite{shif}. As we
said above it is not clear to us how much of the dynamics can be built
over this condensate, and we can let the gluon mass as a free parameter in
Eq.(\ref{e20}). However, we would like to show that this identity is natural.
There are several ways to verify that the gluon mass (or the condensate field)
are related to the phenomenological gluon condensate. We obviously expect that
the dynamical gluon mass is connected to the gluon condensate through the
operator product expansion as described by Eq.(\ref{e13}). We can also compute
the vacuum energy as a function of the propagator in Eq.(\ref{e20}), and than
equalize this vacuum energy to the trace anomaly expression $\frac{\beta
(g)}{2g}  \< G_{\mu\nu} G^{\mu\nu} \>$, where $\beta (g)$ is the perturbative
$\beta$ function of the renormalization group equation. This approach was used
in the last paper of Ref.~\cite{corn} and more recently in Ref.~\cite{mont},
where the relation with the gluon condensate was explored to fix the
gluon mass. As these methods have already been discussed for massive
gluons, we propose here other procedures where we can show that the
gluon propagator of Eq.(\ref{e20}) can indeed be related to the
gluon condensate as well as to other typical hadronic parameters. We start
by showing the relation between the confining interaction and the bag
constant. 
 
The bag constant (${\cal B}$) for a given gluon
propagator can be obtained through~\cite{caro}
\be
{\cal B} = 12\pi^2 \int \frac{k^2 dk^2}{(2\pi)^4} 
\left(
\ln{\left(\frac{A^2(k^2)k^2+B^2(k^2)}{A^2(k^2)k^2}\right)}-
\frac{B^2(k^2)}{A^2(k^2)k^2+B^2(k^2)}\right),
\label{e21} 
\ee
where $A(k^2)$ and $B(k^2)$ appear in the inverse of the renormalized quark
propagator
\be
S^{-1}(k)= \imath\not\!{k} + \Sigma (k) = \imath\not\!{k} A(k^2) + B(k^2). 
\label{e22}
\ee
The form of $A(k^2)$ and $B(k^2)$ is obtained solving the Schwinger-Dyson
equations for a given gluon propagator, and here enter the information about
the GCM propagators.

The Schwinger-Dyson equations for a gluon propagator equal to Eq.(\ref{e11})
were discussed in Ref.~\cite{nat}. To avoid a numerical calculation that
may conceal the simplicity of the problem, we will approximate the
gluon propagator only by its confining part. The part proportional
to $(k^2 + m^2_g)^{-1}$ is responsible for the tail of the propagator,
actually, the larger is the gluon mass the smaller is the contribution
to this tail. Therefore, we will not introduce a large error in neglecting the
second term of Eq.(\ref{e11}) or Eq.(\ref{e20}). Proceeding in this way it is
easy to verify that the propagator $\Delta (k^2) = 16 \pi^4  m_g^2\delta^4(k)$
imply in the following solution for $A(k^2)$ and $B(k^2)$
when $k^2 < m_g^2$~\cite{robwil}
\be
A(k^2) = 2  \,\,\,\,\,\;\; , \,\,\,\,\;\; B(k^2) =  2 \sqrt{m_g^2 - k^2} .
\label{e23}
\ee
For $k^2 > m_g^2$ we have $B(k^2) = 0$. With the solution of Eq.(\ref{e23})
the calculation of ${\cal B}$ is straightforward and gives
\be
{\cal B} = \frac{m_g^4}{16 \pi^2}.
\label{e24}
\ee
For a gluon mass of approximately $600 \, MeV$~\cite{nat,corn} we obtain 
${\cal B} =  (169 \, MeV)^4$ which, considering the approximation
performed to obtain Eq.(\ref{e23}), is in good
agreement with the MIT value of $(146 \, MeV)^4$.

The bag constant can be related to the string tension for fermions in the
fundamental representation through~\cite{kj}
\be
K_F = ( 8 \pi \a_s C_F {\cal B})^{1/2},
\label{e25}
\ee
where $C_F$ is the quadratic Casimir operator for the fundamental
representation. On the other hand this same string tension has
been estimated in the last paper of Ref.\cite{corn} as
\be
K_F \approx \frac{\pi^3}{9} \,\, \frac{\gc}{m_g^2}.
\label{e26}
\ee
From the identity of Eq.(\ref{e25}) and Eq.(\ref{e26}) we obtain
\be
m_g^4 = \frac{\pi^4}{9} \left( \frac{3}{2\pi \a_s} \right)^{1/2} \gc .
\label{e27}
\ee
Eq.(\ref{e27}) shows the connection between the gluon mass and the
gluon condensate. Taking $\a_s \approx 1$ and 
$\gc \simeq (0.01) \,\, GeV^4 $~\cite{shif} we obtain
$m_g \approx 550 \,\, MeV$, which is consistent with the
value we quoted before (see Refs.~\cite{nat,celenza}).

Another way to verify the relation between the gluon mass and the
gluon condensate is comparing the bag energy density to minus the
vacuum expectation value of the trace of
the QCD energy momentum tensor, although it is known that
this comparison does not give reasonable numbers~\cite{shuryak}. The energy
density in the bag model at the equilibrium radius is equal to
\be
\Omega = 4 {\cal B},
\label{e28}
\ee
and from the vacuum expectation value of the trace of
the QCD energy momentum tensor we obtain 
\be
\Omega = \frac{1}{4} \< \Theta_{\mu\mu} \> = - \frac{1}{24} \frac{\a_s}{\pi} (11N-2n_f)
\< G_{\mu\nu} G^{\mu\nu} \>.
\label{e29}
\ee
which gives
\be
m_g^4 = \frac{(11N-2n_f)\pi^2}{6} \gc.
\label{e30}
\ee
As expected we obtain a gluon mass proportional to the gluon condensate,
but about $1.5$ times the desired value (considering $n_f=3$).

In conclusion, we discussed how the QCD vacuum model of Celenza and
Shakin~\cite{celenza2} explain the presence of a confining gluon propagator
frequently used in calculations within the global colour model.
In particular, it reproduces the propagator of Ref.~\cite{nat}
where the dynamically generated gluon mass plays a fundamental role.
We considered the strength of the confining interaction, equal to the
gluon mass, to be a free parameter contrarily to Ref.\cite{celenza},
based on the fact that the dynamics may modify the gluon propagator.
We have given arguments showing that the gluon mass is indeed
related to the phenomenological gluon condensate, as well as it can be
related to the string tension and bag constant.
The coincidence between the propagators used
in applications of the global colour model and the one resulting
from the squeezed vacuum model of Celenza and Shakin might
indicate some intrinsic property of the actual QCD vacuum.

\section*{Acknowledgments}

I am grateful to A.~E.~Dorokhov for valuable remarks.
This research was supported in part by the Conselho Nacional
de Desenvolvimento Cientifico e Tecnol\'ogico (CNPq), 
Funda\c c\~ao de Amparo a Pesquisa do
Estado de S\~ao Paulo (FAPESP), and by Programa de Apoio a N\'ucleos
de Excel\^encia (PRONEX).


\end{document}